**Low Field, Current-Hysteretic Oscillations in Spin Transfer Nanocontacts**

M. R. Pufall, W. H . Rippard, M. Schneider, S. E. Russek

*Electromagnetics Division, National Institute of Standards and Technology, Boulder, CO 80305*

## Abstract

We have measured spin-transfer-driven, large amplitude, current hysteretic, low frequency (< 500 MHz), narrowband oscillations in nanocontacts made to spin valve structures. The oscillations occur in zero field, persist up to 5 mT for in plane applied fields, and to beyond 400 mT for out of plane fields. Unlike previous measurements, the oscillation frequency is well below that for uniform-mode ferromagnetic resonance, is only a weak function of applied field, and is highly anharmonic. The oscillations are hysteretic with applied dc current, appearing at high currents but persisting to lower currents upon decrease of the current. We suggest that these observations are consistent with the dynamics of a nonuniform magnetic state in the vicinity of the contact nucleated by both the spin transfer torque and dc current-generated Oersted fields, with the dynamics driven by spin transfer. The electrical oscillation amplitudes are large and narrowband, with the largest amplitudes on the order of 1 mV and the narrowest linewidths below 1 MHz.

Since the prediction that spin-polarized currents can exert significant torques in magnetic nanostructures, a wide variety of magnetization dynamics driven by spin transfer torques have been observed in a wide range of device geometries and experimental conditions.[1,2] The general characteristics of these observed dynamics—the amplitude, the fundamental excitation frequency $f_0$, the change of $f_0$ with current $I$ and applied field $\mu_0 H_{app}$—are roughly understandable using theories that approximate the free layer dynamics as quasi-uniform large angle magnetization motion in the region of the device where current flows.[3,4] In the case of nanopillars, this region is the entirety of the free layer (possibly ignoring some region at the edge), and in nanocontacts consists of the region directly under the contact; in the latter case the mode remains centered (*i.e.*, stationary) on the symmetry axis. Even this rough correspondence between theory and experiment is somewhat surprising, since one might expect excitations with nonuniform magnetization (on the scale of the contact) due to the large spatially-varying Oersted fields generated by the dc current itself. The effect of these fields, which approach 6.5 mT/mA (65 Oe/mA) at the edge of a 60 nm diameter contact, is an active area of computational magnetic research.[5] In this Communication we present measurements of large amplitude, narrowband signals from nanocontacts that are *not* easily explainable using such radially symmetric quasi-uniform mode approximations. The measurements were performed on nanocontacts nominally identical to those measured previously. The principal difference in the results reported here is that the in plane field magnitude is always less than 5 mT, whereas previously this magnitude was greater than 60 mT. We suggest that the observed dynamics may result from the generation and perturbation of a nonuniform magnetic state, such as a magnetic vortex, in the vicinity of the contact.

We observe oscillations with frequencies less than 500 MHz that are present only from zero to 5 mT for in plane fields, and persist to above 0.4 T for fields applied directly out of plane. For either field orientation, the oscillation frequency is typically significantly below the uniform-mode ferromagnetic resonance (FMR), and changes very little with applied field strength. Finally, the *presence* of the oscillations is hysteretic with dc current, with oscillations appearing at a high current with increasing current but persisting to lower currents upon decrease of the current. These results are markedly different from results presented previously,[6,7] which showed significant $df_0/dI$ and $df_0/dH_{app}$, higher frequencies, and no current hysteresis. The field dependence in particular is consistent with measurements of vortex dynamics in patterned microstructures.[8, 9] Furthermore, the current-generated Oersted fields are among the largest applied fields in the system, suggesting that the hysteresis observed may be due to the nucleation of a nonuniform mode by the combination of Oersted fields and spin transfer, the dynamics of which are also stabilized by the circumferential Oersted field.

The devices presented here are pseudo-spin valves comprising Ta (3 nm)/Cu (15 nm)/$Co_{90}Fe_{10}$(20 nm)/Cu (4 nm)/$Ni_{80}Fe_{20}$(5 nm), essentially similar to devices measured previously in high fields and discussed in detail elsewhere.[6,7] The oscillations were observed for a range of contact resistances. Little correlation was observed between resistance and oscillation characteristics, other than that high resistance (> 15 Ω) contacts tended to show oscillations less frequently than low resistance contacts. We also observed similar low frequency excitations in other material systems, but these will not be discussed here. Note that the fixed layer is unpinned, raising the possibility that this layer is also modified by the Oersted fields and is involved in the dynamics.[10] The spin valve was patterned into a 10 μm x 20 μm mesa and a contact with 60 nm - 80 nm nominal diameter made to the multilayer via an e-beam lithography process.[6] The devices were measured at room temperature, with magnetization dynamics detected electrically via the giant magnetoresistance (GMR) effect [11] and measured by either a spectrum analyzer or a real-time oscilloscope with 1.5 GHz ($8x10^9$ samples per second) bandwidth.

Typical results for in plane applied fields are shown in Fig. 1. In the measurement, a large saturating field is first applied, and the field then reduced to the measurement value. The dc current through the contact is ramped from 4 mA to 12 mA and back down to 4 mA. In this way, hysteresis in the output vs. dc current can be identified. Figs. 1a and 1b show contour plots of the spectral output vs. current for 1 mT and 3 mT, respectively. In Fig. 1a, the device produces no ac output until just below 11 mA—the "turn-on" current—whereupon it emits a signal at approximately 188 MHz along with strong second and third harmonics. This narrowband signal (18 MHz full width at half maximum; note the log scale in the plot) persists with slightly increasing frequency until just below 12 mA, where the output evolves into a signal with lower power density. This evolution is typical for the measurements presented here, with narrowband signals evolving into a broader-band output at larger currents, possibly due to nonlinearities driven by the increasing spin transfer torque.

Upon decrease of the current from 12 mA, the large amplitude narrowband signal re-emerges, but persists as the current drops below 11 mA, the turn-on current. The linewidth narrows as the current decreases, reaching a minimum (for this particular device and geometry) of approximately 4 MHz at 8 mA, and the frequency decreases at roughly 10 MHz/mA. The output ceases just above 4 mA—the "turn-off" current—demonstrating substantial hysteresis in the presence (but not frequency) of oscillations with current. This hysteretic behavior has been observed in many devices, with the particulars—onset current, fundamental frequency, relative harmonic power, and hysteresis—varying from device to device.[12]

Increasing the in plane field alters the range of currents over which such output is observed, but does not appreciably change the output frequency itself. For an applied field of 3 mT (Fig. 1b) the turn-on current decreases below that observed for 1 mT, while the turn-off current increases to above 6 mA. The turn-on and turn-off currents tend to approach each other with increasing field, and the narrowband oscillations cease above a certain field (about 5 mT for the device in Fig.1), leaving only the lower power density, broadband output seen at higher currents. The output frequencies remain relatively constant from zero field through 4 mT, as shown in Fig. 1c, with significant variations seen primarily near the turn-off current. Finally, as shown in Fig. 1d, the integrated power of the fundamental frequency is insensitive to applied fields, further indicating that the mode is not strongly affected by $H_{app}$. Interestingly, reversing the in plane field direction gives very similar oscillation frequencies, but does not yield identical results for the current hysteresis, relative harmonic power, and field dependence. This may indicate that local effective field variations in these films are of similar magnitude to these applied fields.

The results presented here are markedly different from those reported previously for in plane fields greater than 50 mT[6], which showed a higher frequency mode that redshifted with current ($df_0/dI \approx$ -200 MHz/mA), and an appreciable $df_0/d\mu_0 H$ (≈ 25

GHz/T). By contrast, the results presented here have $\mathrm{d}f_0/\mathrm{d}I \approx +10$ MHz/mA and $\mathrm{d}f_0/\mathrm{d}\ \mu_0 H \approx 0$. Also, the measured output power at $f_0$ in the higher field oscillations was typically more than an order of magnitude smaller than that presented here. Most significantly, the higher field oscillations did not show hysteresis with dc current as seen here. While the previously published results were roughly understandable as large-angle versions of quasi-uniform mode precession, we see no such correspondence for the results presented here. For example, the theoretical uniform mode ferromagnetic resonance (FMR) frequency for a NiFe film with 0.5 mT in plane uniaxial anisotropy increases from 600 MHz to 2.1 GHz from zero to 5 mT for in plane fields, whereas here the frequency is effectively constant. Large-angle in-plane uniform precession also predicts that $\mathrm{d}f_0/\mathrm{d}I < 0$, at least near the critical current.[4,6]

When the applied field is directed along the surface normal, these oscillations persist in much larger applied fields than for in plane fields. As shown in Fig. 2a, 20 mT applied out of plane (to the same device presented in Fig. 1) does not suppress the narrowband output. The frequency of this mode is again a weak function of current, increasing at about 8 MHz/mA. Out of plane fields also reduce the current-induced hysteresis (as did in plane fields), but without changing the turn-off current significantly. Out of plane fields generally reduce the turn-on current, resulting in a spectral output roughly symmetric in current (see Fig. 2b).

A major effect of larger out of plane fields is to decrease the onset current for a broader-band, more rapidly blueshifting output, seen at 10 mA in Fig. 2a, 9 mA in Fig. 2b. Thus, the current range over which narrowband precession is observed shrinks with increasing field. Above a certain field (in this case about 500 mT) only the broad blueshifting mode remains, along with another broadband signal at higher frequency, seen just appearing at $I$ = 12 mA in Fig. 2b at 1.3 GHz, whose onset current also decreases with field strength. Line plots of the spectral outputs at low and high currents are shown in Fig. 2c for $\mu_0 H_{app}$ = 300 mT, showing narrowband multiharmonic output at low currents and the two broader modes that develop at higher currents. Larger fields eventually suppress these low frequency outputs so that the broadband gigahertz frequency oscillations presented here do not connect at higher fields with the narrowband high frequency oscillations reported previously.[6] As shown in Fig. 2d, the frequency of the narrowband output is again only a weak function of out of plane field strength, initially decreasing with field at -150 MHz/T below 140 mT, and then increasing at a similar rate above this field.

These results are again different from those observed previously,[6] which showed that for large (> 600 mT) out of plane fields, the observed precessional mode was not hysteretic in current, and the frequency was a strong function of both current and field, increasing with field at 30 GHz/T. Those results roughly followed that expected for large-angle uniform mode precession (but with some significant unexplained differences).[7,13] The current results also differ from FMR, for which the calculated small-angle FMR decreases monotonically from 600 MHz to 400 MHz from zero to 400 mT for out of plane applied fields.

The power of the fundamental frequency and its harmonics is typically substantial for this low frequency mode, as shown in Fig. 3 for a different device. Fig. 3a shows the spectral output from 14 mA to 8 mA, while Fig. 3b shows the spectrum at 11.75 mA. The peak-to-peak amplitude of the fundamental (unamplified, and uncorrected for device impedance mismatches) approaches 0.3 mV for this device; other devices show amplitudes over 1 mV. For comparison, the amplitude for full ($180^{o}$) precession should be 2 mV to 4 mV, inferred from the dc GMR of other nanocontacts. At the same time, the integrated powers of the harmonics sum to approximately 60 % of the power in the fundamental frequency. Interestingly, in this case the linewidth of the fundamental is 575 kHz, the lowest observed thus far in a nanocontact made to a NiFe-CoFe spin valve, in any field geometry. This may be an intrinsic property of this mode (*e.g.*, somehow having greater thermal stability) and may also be related to the small $df_0/dI$ and $df_0/d\,\mu_0 H$, which reduce noise modulation broadening.

A single-shot measurement of the output voltage vs. time from this device is shown in Fig. 3c, with the oscillator in the same state as for the spectrum analyzer measurement shown in Fig. 3b. The spectrum in Fig. 3b corresponds to a highly nonsinusoidal waveform. A multi-harmonic sinusoidal fit to the waveform is shown, and gives frequencies and amplitudes in good agreement with those found from Lorentzian fits to the spectrum. The RMS noise at the oscilloscope is 0.35 mV, accounting for some of the variations seen in the time trace.

The frequency of the fundamental and its weak current and field dependencies indicate that this low frequency mode is substantially different from previously measured (quasi-uniform mode) spin transfer resonances, and also different from FMR. Other observations presented above—the current-induced hysteresis in the ac output, fundamental amplitude, and the harmonic power—also suggest this. This suggests that instead of quasi-uniform mode dynamics, a *non*stationary mode with *non*uniform magnetization (on the scale of the contact) is being driven in these measurements. While the magnetization distribution of this system on the length scale of the contact is unknown at this point, one process that reasonably accounts for these results is the nucleation of a nonuniform vortex-type state by the dc current, through a combination of the spin transfer torque and the Oersted fields generated by the current itself.

In patterned magnetic structures such as a micrometer-sized disk, in which a vortex is the lowest-energy state, the vortex will circulate about the structure's center in response to a perturbation.[8-10,14] The restoring force is provided by demagnetizing fields due to the finite device size, with the frequency scaling as the disk thickness and inversely with disk diameter. The vortex oscillation frequency is a weak function of applied field, and is annihilated in sufficiently large field, the magnitudes of which differ with field orientation. The main points of correspondence between the data presented here and vortex dynamics are the weak dependence of frequency on field, the differing quenching field for in- and out of plane fields, and the hysteresis with dc current.

While vortices in typical patterned structures (diameters on the order of micrometers, thicknesses on the order of 20 nm) show frequencies similar to those reported here, the magnetics of the two systems are clearly quite different. In the nanocontact geometry, the lack of a boundary requires that the circumferential Oersted field (or interlayer interactions) provide the restoring force. The field magnitude varies to a good approximation as $r$ inside the contact (where $r$ is the radial distance from the center of the contact), and $1/r$ outside.[15] The 75 mT generated at the edge of a 60 nm contact by a 12 mA current is the largest applied field in the system for many of the results presented above. Micromagnetic simulations[16] indicate that such fields alone (*i.e.*, without spin transfer effects) are sufficient to nucleate and stabilize a vortex-like state at the contact in one or both layers, albeit in mesas smaller than those measured here. These simulations further show that this circumferential field stabilizes an existing vortex at the contact so that once nucleated, the vortex persists to currents below the nucleation current, consistent with the observed hysteresis with current.[17] Also, because the angle of the magnetization varies across a nonuniform state, the oscillation of such a distribution around the contact center would result in effective large-angle magnetization motion, and hence in large output powers via GMR. For example, the magnetization vectors on either side of the core of a vortex are antiparallel.

Though these results are suggestive, a more quantitative comparison is complicated by both the conjectured nonuniform magnetization, and the spin transfer effect. If both the fixed and free layers are affected by the Oersted fields, there may be significant dipolar coupling across the 5 nm spacer because of the stray fields from the nonuniform magnetization. Beyond this, the spin transfer torque for such currents drives large angle magnetization dynamics in other geometries,[6,18] and so will likely also play a large role here. Other work has measured the anisotropic magnetoresistance and spin transfer excitation of vortices in nanostructures with a single magnetic layer and current in the plane of the film.[19] In the present case with current perpendicular to the plane, nonuniform magnetization in the contact region of both the free and fixed layers makes determining the role of spin transfer torque in driving these dynamics a challenge, and an area of ongoing study.[20]

We have shown that low frequency (less than 500 MHz), current-hysteretic dynamics are excited in spin transfer nanocontacts. These oscillations are only observed for zero and small (< 5 mT) in plane fields, but can occur in larger (< 500 mT) out of plane fields. The frequency of oscillation is a very weak function of field for either geometry. The frequency peak is also a weakly blueshifting function of dc current, and its presence/absence is a hysteretic function of current, appearing at high currents and persisting to lower currents.  The precession typically has a large harmonic content, indicating significant eccentricity in the relative motion of $<\boldsymbol{M}_{\mathrm{free}}>$ and $<\boldsymbol{M}_{\mathrm{fixed}}>$. These oscillations frequently exhibit linewidths less than 1 MHz, and amplitudes approaching 1 mV, some of the smallest and largest such values, respectively, observed in spin transfer driven systems, and as a consequence warrant further study.

**Endnotes:**

[1] J.A. Katine, F. J. Albert, R.A. Buhrman, E.B. Myers, D. C. Ralph, Phys. Rev. Lett. **84,** 3149 (2000); B. Ozyilmaz *et al.*, Phys. Rev. Lett. **91**, 067203 (2003); J. Grollier *et al.*, Phys Rev. B **67**, 174402 (2003).

[2] M. Tsoi *et al.*, Nature **406,** 46 (2000); S. Urazhdin, N. O. Birge, W. P. Pratt, J. Bass, Phys. Rev. Lett. **91,** 146803 (2003); M. Covington, *et al.*, Phys. Rev. B **69,** 184406 (2004); M. R. Pufall *et al.*, Phys. Rev. B **69,** 214409 (2004).

[3] J. C. Slonczewski, J. Magn. Magn. Mater. **159,** L1 (1996).

[4] S. M. Rezende, F. M. de Aguiar, A. Azevedo, Phys. Rev. B **73,** 094402 (2006); M. A. Hoefer, M. J. Ablowitz, B. Ilan, M. R. Pufall, T. J. Silva, Phys. Rev. Lett. **95,** 267206 (2005); A. N. Slavin, V. S. Tiberkevich, Phys. Rev. Lett. **95,** 237201 (2005).

[5] D.V. Berkov and N. L. Gorn, J. Appl. Phys. **99** (8), 08Q701 (2006).

[6] W. H. Rippard, M. R. Pufall, S. Kaka, S. E. Russek, T. J. Silva, Phys. Rev. Lett. **92,** 027201 (2004).

[7] W. H. Rippard, M. R. Pufall, S. Kaka, S. E. Russek, T. J. Silva, Phys. Rev. B **70** (10), 100406 (2004).

[8] V. Novosad et al., Phys Rev. B **72**, 024455, (2005); K. S. Buchanan *et al.*, Phys. Rev. B **74**, 064404 (2006).

[9] K. Yu. Guslienko *et al.*, J. Appl. Phys. **91**, 8037 (2002).

[10] K.W. Chou *et al.*, J. Appl. Phys **99**, 08F305 (2006).

[11] M. R. Pufall, W. H. Rippard, S. E. Russek, S. Kaka, J. A. Katine, Phys. Rev. Lett. **97** (8), 087206 (2006).

[12] We have not yet quantified by how much these quantities vary across devices, but compare Fig. 1 to Fig. 3 for an example.

[13] W. H. Rippard *et. al*, accepted for publication in Phys. Rev. B.

[14] J. P. Park and P. A. Crowell, Phys. Rev. Lett. **95**, 167201 (2005); S.-B. Choe *et al.*, Science **304**, 420 (2004).

[15] M. A. Hoefer and T. J. Silva, lanl.arXiv.org/cond-mat/0609030.

[16] Implemented in open source OOMMF: http://math.nist.gov/oommf/software.html

[17] A simple calculation of the frequency of small oscillations based on a single rigid vortex model (Ref. 9) overestimates the fundamental frequency (≈ 1 GHz at 1 mA) and predicts a linear dependence of frequency on current, quite different from our observations. It does, however, indicate that the Oersted field exerts a substantial restoring force on a vortex.

[18] S.I. Kiselev *et al.,* Nature **425**, 380 (2003).

[19] P. Vavassori et.al, Appl. Phys. Lett. **86**, 072507 (2005); S. Kasai *et al.*, Phys. Rev. Lett. **97**, 107204 (2006).

[20] For example, currents of opposite sign, *i.e.*, *negative* currents, sometimes—but infrequently—drive small-angle dynamics for low fields. These are never observed for high fields, and should occur only if 1) $\boldsymbol{M}_{\text{free}}$ and $\boldsymbol{M}_{\text{fixed}}$ are not parallel, and 2) the parallel configuration is somehow destabilized. This would occur if one, but not both, of the layers were in a vortex-like state.

## Figure Captions

**FIG.1:** (Color Online) Output from nanocontact for in plane fields. **a)** Contour plot of spectral output for 1 mT (10 Oe). Note symmetric current scale: current is ramped up to 12 mA, then down. Color denotes power, logarithmic scale to show low power harmonics; **b)** Contour plot of spectral output, 3 mT. Results of Lorentzian fits to spectra: **c)** fundamental frequency vs. $I$, for several fields; **d)** Power of fundamental vs. $I$, for several fields. Both up and down scans of $I$ are plotted.

**FIG. 2:** (Color Online) Output for out of plane fields. **a)** Contour plot of spectral output for 20 mT. Color again denotes power, logarithmic scale **b)** Contour plot of spectral output, 200 mT. **c)** Spectral output at two currents, $\mu_0 H_{app}$ = 300 mT. **d)** Fits to fundamental frequency (highest power) vs. $I$ for several fields.

**FIG. 3:** (Color Online) Comparison of spectral and temporal outputs. Amplitudes include 30 dB amplification. **a)** Spectral output from device vs. dc current, $\mu_0 H_{app}$ = 2.5 mT; **b)** Log-power spectrum at I = 11.75 mA, showing harmonics. Fitted $V_{fund}$ = 1.94 mV; **c)** Time trace of device output at I = 11.75 mA and fit of fundamental and 3 harmonics. Fitted $V_{fund}$ = 1.99 mV.

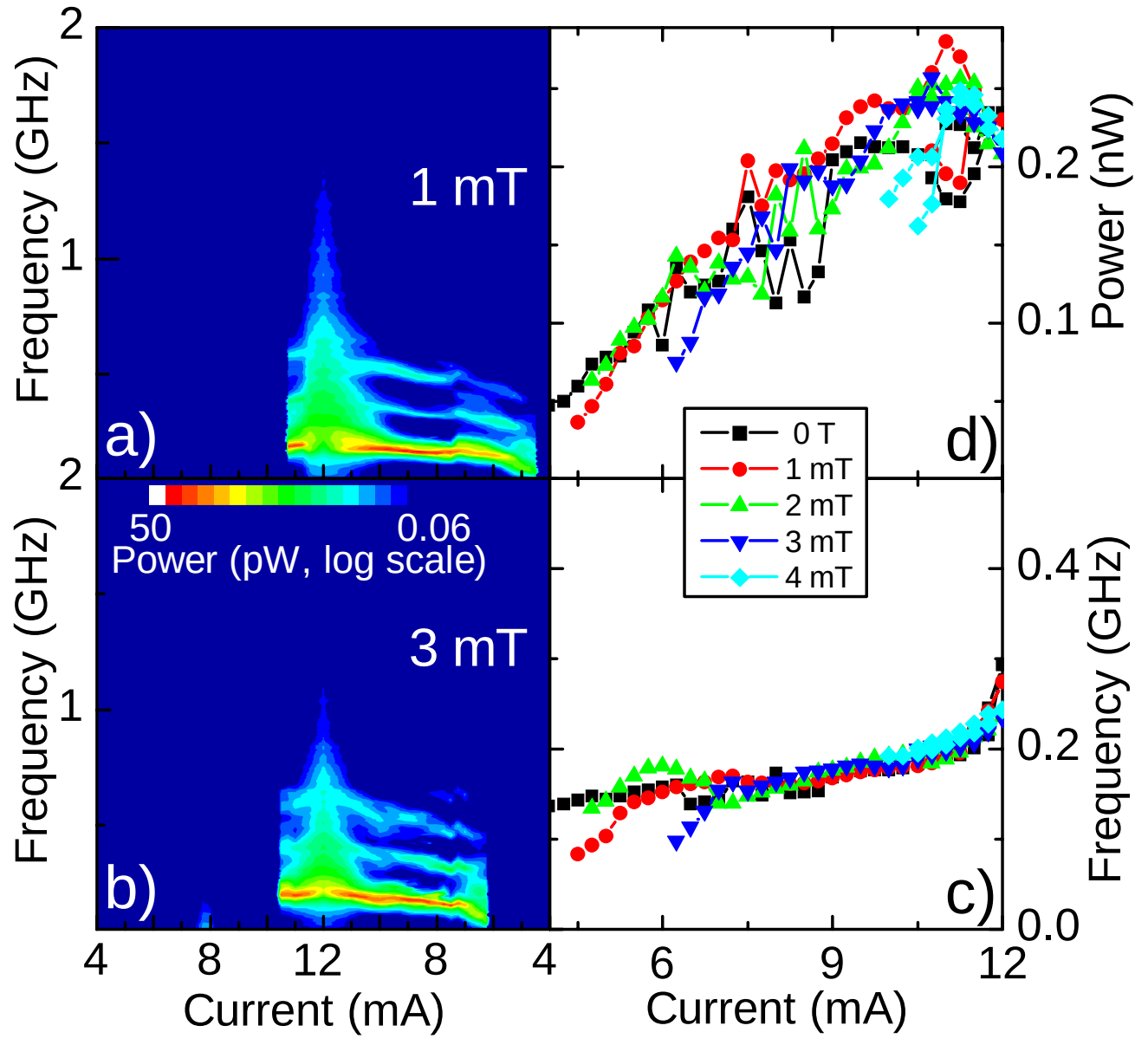

**Figure 1**

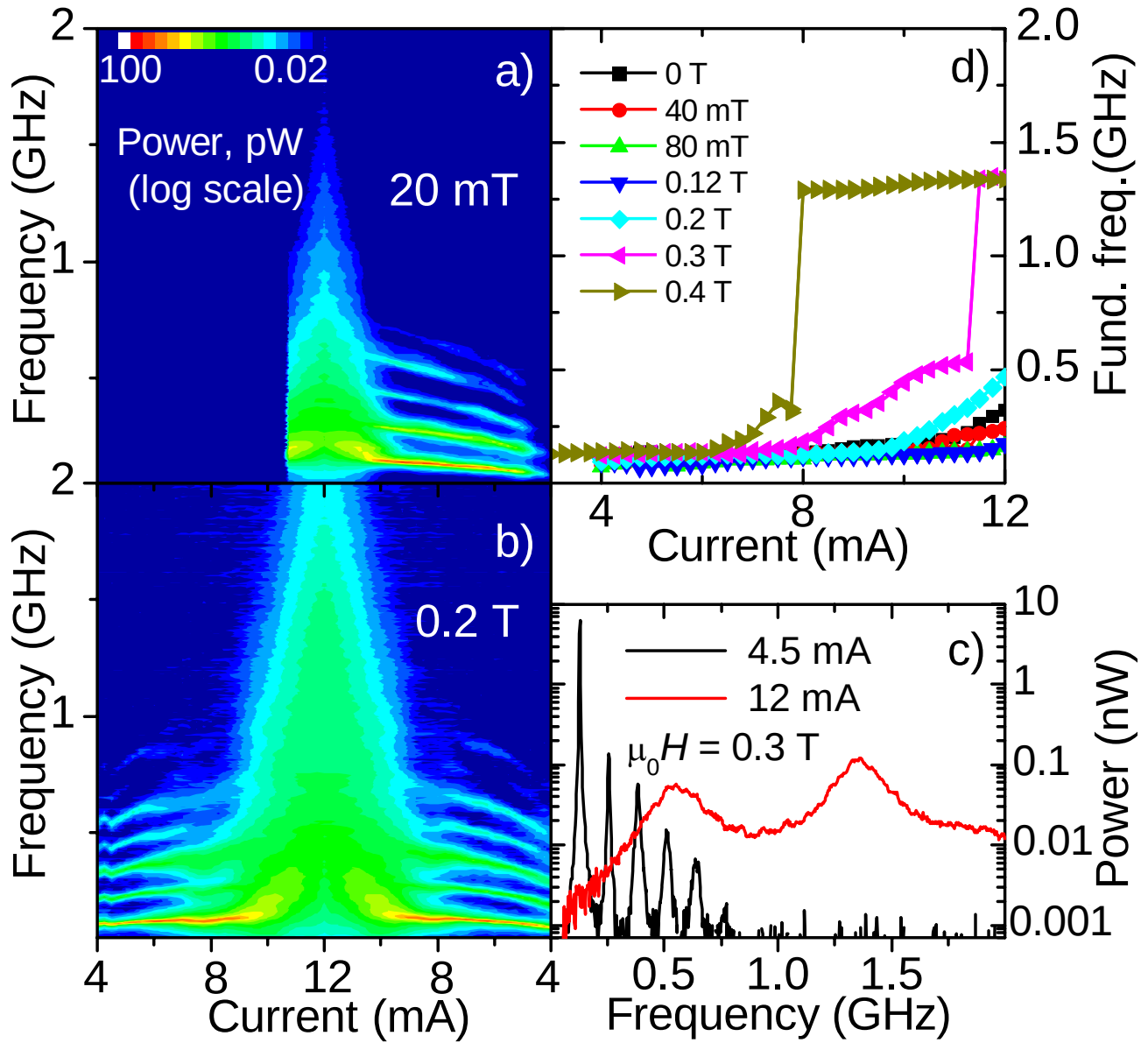

**Figure 2**

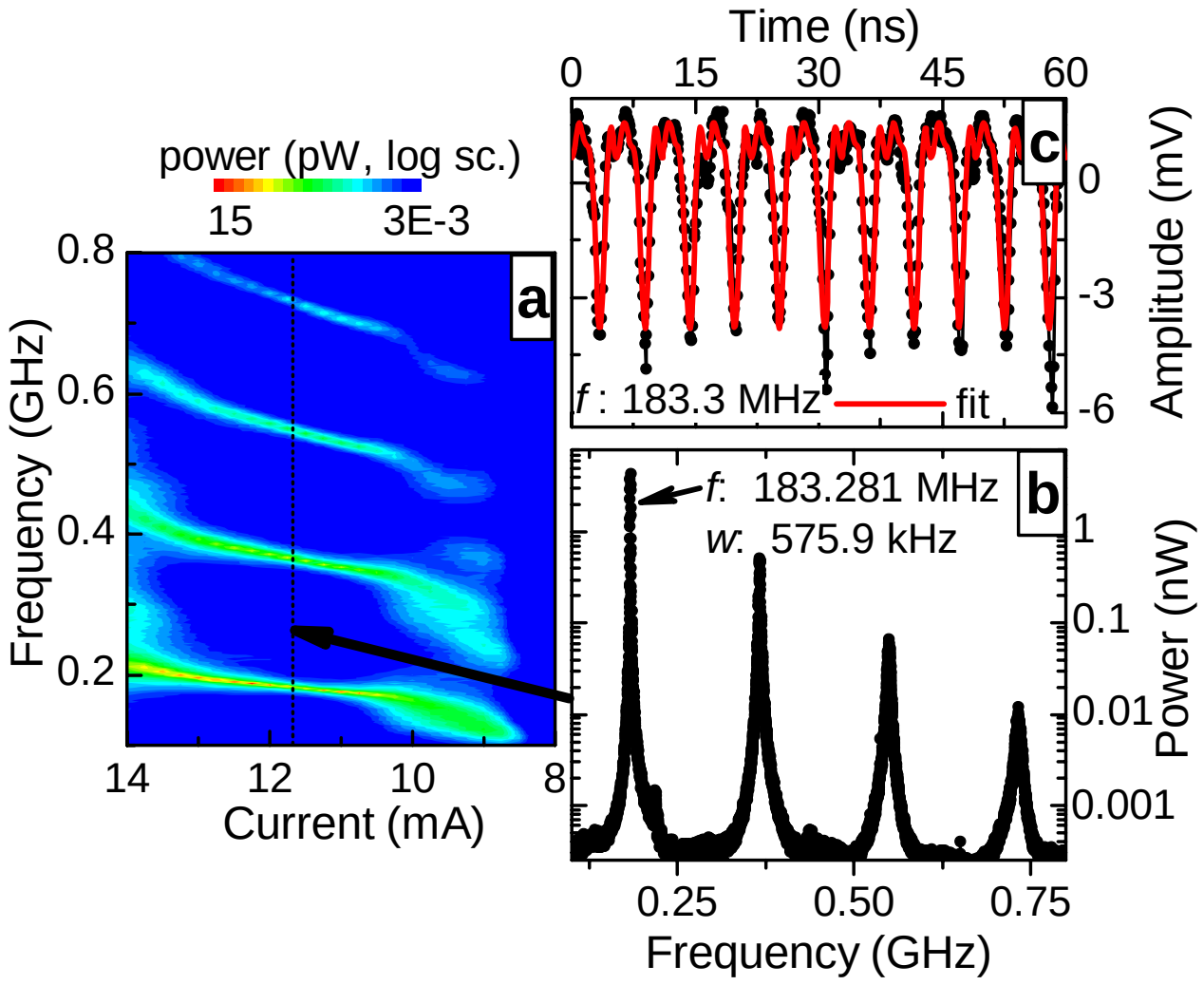

**Figure 3**